\documentclass[aps,preprintnumbers,twocolumn,showpacs ]{revtex4}
\usepackage{amsmath,amssymb,graphicx}
\usepackage{epsfig}
\usepackage{graphicx}
\usepackage{enumerate}
\usepackage{graphicx,wrapfig}

\usepackage{color}

\makeatletter
\@ifundefined{textcolor}{}
{%
 \definecolor{BLACK}{gray}{0}
 \definecolor{WHITE}{gray}{1}
 \definecolor{RED}{rgb}{1,0,0}
 \definecolor{GREEN}{rgb}{0,1,0}
 \definecolor{BLUE}{rgb}{0,0,1}
 \definecolor{CYAN}{cmyk}{1,0,0,0}
 \definecolor{MAGENTA}{cmyk}{0,1,0,0}
 \definecolor{YELLOW}{cmyk}{0,0,1,0}
 }

\newcommand{\bea}{\begin{eqnarray}}
\newcommand{\eea}{\end{eqnarray}}

\newcommand{\be}{\begin{equation}}
\newcommand{\ee}{\end{equation}}

\begin{document}

\title
{Towards equilibration and thermalization between finite quantum systems:
The role of dephasing effects and inelastic interactions}

\author{Manas Kulkarni$^{1,2}$, Kunal L. Tiwari$^{1,2}$, Dvira Segal$^1$ }
\affiliation{$^1$Chemical Physics Theory Group, Department of Chemistry,
University of Toronto, 80 Saint George St. Toronto, Ontario, Canada
M5S 3H6}
\affiliation{$^2$ Department of Physics, University of Toronto, 60 Saint George St. Toronto, Ontario, Canada
M5S 1A7}


\date{\today}

\begin{abstract}
We demonstrate the approach towards a Gibbs-like equilibrium state,
with a common temperature and a chemical potential, of two finite
metallic grains, prepared with a different number of
noninteracting electrons, connected by a weak link that is
susceptible to incoherent and inelastic processes. By developing an analytic method
and by using an exact numerical approach, the quantum time evolution
of the electrons in the metallic grains is followed. In the absence of
decoherring and inelastic effects, equilibration
is never reached. Introducing dephasing effects on the link {\it
only}, using a dephasing probe, the two quantum systems equilibrate,
but do not evolve towards a Gibbs-like state. In contrast, by
mimicking inelastic interactions with a voltage probe, the metal
pieces evolve towards a common Gibbs-like equilibrium state, with
the probe.
\end{abstract}


\maketitle


{\it Introduction.} How do quantum systems equilibrate from a
certain non-equilibrium initial condition, e.g., a quench
\cite{Rev}? With cold atoms in optical lattices offering a clean
experimental setup \cite{Cold}, renewed attention in this
fundamental topic has recently sparked. One could address this
question with (at least) three distinct setups in mind: (i) Consider
an {\it isolated} quantum system, and study its evolution towards
equilibrium, for example, in the time averaged sense
\cite{Reimann,Short}. (ii) Attach a subsystem with a few degrees
freedom to a thermal reservoir, and monitor the system
equilibration,
 e.g., in the sense of the small trace distance
\cite{Popescu,Trace}, or (iii) put in contact two identical {\it
finite} quantum systems and watch for the process of mutual
equilibration \cite{Hanggi}. Despite intense 
efforts, general results are still missing \cite{Yukalov}.  Recent studies also argue
about the precise definition of quantum integrability, and its implication on
quantum thermalization  \cite{Integ1,Integ2}.

In this work, we consider a combined setup
and demonstrate the process of mutual equilibration
of two finite metallic quantum systems, connected by a weak link, in
the presence of either elastic-decoherring processes or inelastic
effects, through the interaction of the link electrons with
additional degrees of freedom.
When only decoherence effects are
allowed, the system approaches a non-canonical equilibrium state. In
contrast, when inelastic processes are included, the two parts relax
towards a common Gibbs-like state.
The origin of inelastic scattering processes are many-body
interactions in the system, e.g., coupling electrons to phonons.
Since an explicit and exact inclusion of such effects is challenging
\cite{Eran,Thoss,Thorwart,Andrei}, it was suggested 
\cite{Buttiker,Beenakker} to phenomenologically introduce elastic
and inelastic scattering processes by using dephasing or voltage
probes, respectively. These electron reservoirs are prepared such
that there is no net (energy resolved or total) electron flow from
the system towards these probes.

\begin{figure}[htbp]
\hspace{4mm} {\hbox{\epsfxsize=50mm \epsffile{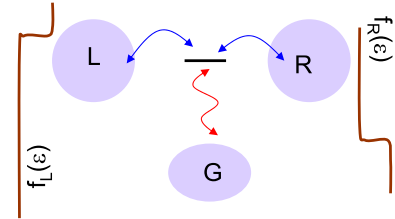}}} \caption{Two
metallic grains (reservoirs) separately prepared in a grand canonical -
diagonal state; the initial population is plotted at the
boundaries. At $t_0$ the reservoirs are put into contact through a
single electronic state, susceptible to decoherring and relaxation
effects.
This is accounted for by the $G$ reservoir, serving as
a dephasing or a voltage probe.
}
 \label{FigS}
\end{figure}

As a particular realization, we consider the non-interacting Anderson model with a
single electronic level (dot) coupled to two metallic grains
(henceforth referred to as reservoirs) \cite{Anderson}. Each
reservoir is initially prepared in a distinct Gibbs-like grand
canonical state, at a different chemical potential. We follow the
time evolution of the reservoirs' electrons, once put in contact
through the dot, itself susceptible to decoherring and/or inelastic
processes. For a schematic representation, see Fig \ref{FigS}. We
refer to the metal grains, including (each) $N\sim 500$ electronic
states and $n\sim 200$ electrons as ``reservoirs", to indicate that
they have a dense-enough density of states, such that their effect
on the impurity (dot) can be absorbed into a positive real
self-energy function, allowing for a quantum Langevin equation (QLE)
description \cite{QLE}.
While we may consider large reservoirs,  the number of electrons
in the metal grains-dot unit is fixed. However, energy is conserved only in the elastic scattering
scenario.


{\it Model.} We consider two electron reservoirs $\nu=L,R$, with
identical density of states and a sharp cutoff at $\pm D$,
which do not directly couple, only through their weak hybridization with a
single-state quantum dot. The Hamiltonian takes the form
\begin{equation}
H_0=H_{L}+H_{R}+H_{W}+V_{L}+V_{R},
\end{equation}
where $H_{L,R,W}$ represents the Hamiltonian for left reservoir,
right reservoir and dot, respectively. 
 The term $V_{\nu}$ denotes the coupling of
the dot to the $\nu$ reservoir,
\begin{eqnarray}
H_{L}&=&\sum_{l}\epsilon_{l}c_{l}^{\dagger}c_{l},\,\,\,
H_{R}=\sum_{r}\epsilon_{r}c_{r}^{\dagger}c_{r}, \,\,\, H_{W}=\epsilon_{d}c_{d}^{\dagger}c_{d}
\nonumber\\
{{V_{L}}}&=&\sum_{l}v_{l}c_{d}^{\dagger}c_{l}+h.c. \,\,\,\,
{{V_{R}}} =\sum_{r}v_{r}c_{d}^{\dagger}c_{r}+h.c. \label{eq:H0}
\end{eqnarray}
Here, $c^{\dagger}_{k}$ ($c_k$) are fermionic creation
(annihilation) operators of the left reservoir, $l\in L$, right
reservoir, $r\in R$ or the dot ($d$). We assume that $v_l$ and $v_r$ are
real numbers and that the Hamiltonians of the metal grains are diagonal in
momentum basis. A factorized initial state is assumed, with an empty
dot and the reservoirs prepared in a diagonal (grand canonical) state at
a chemical potential $\mu_{\nu}$ and inverse temperature
$\beta=1/T$, satisfying the distribution
$f_{\nu}(\epsilon)=[e^{\beta(\epsilon-\mu_{\nu})}+1]^{-1}$,
$\mu_L=-\mu_R$.

\begin{figure}[htbp]
\hspace{2mm}
{\hbox{\epsfxsize=70mm \epsffile{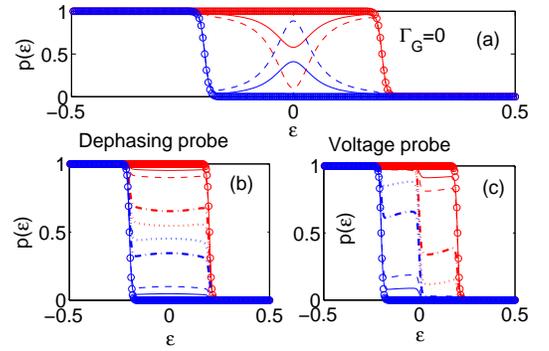}}} \caption{(a) Coherent
population dynamics of the reservoirs' electrons. Plotted are the
$L$ (three top lines) and $R$ (three bottom lines) occupations as a
function of electron energy, for $\Gamma_G=0$ at $t=0$ ($\circ$)
$t=750$ (full) and $t=1500$ (dashed). (b) Approaching a
non-canonical equilibrium state with a dephasing probe,
$L$-bath  (five top lines) and
$R$-bath (five bottom lines) occupation functions.
(c)
Approaching thermal equilibrium with a voltage probe,
$L$-bath (five right-most lines) and
$R$-bath (five left-most lines) occupation functions.
In (b) and (c)
$\Gamma_G=0.4$ and $t=0$ ($\circ$), $t=750$  (full), $t=1500$ (dashed), $t=7500$
(dashed-dotted) and $t=15\times$ $10^3$ (dotted). The last two lines
were generated by restarting the QLE dynamics when approaching
$t=m\tau_{rec}^0$, $m$ is an integer. $\mu_G$ has been further
updated at each restart point to eliminate leakage. In all panels
$\beta=200$, $\Gamma_L=\Gamma_R=0.025$. $\epsilon_d=0$,
$\mu_L=-\mu_R=0.2$, $D=1$, $N=500$  electronic states at each reservoir, $L$, $R$, $G$.} \label{FigG1}
\end{figure}

At $t_0$ the two reservoirs are put into contact through the dot,
and their dynamics is followed using either an exact quantum evolution
scheme, or a QLE approach (details below). The observed population
dynamics is depicted in Fig. \ref{FigG1}(a). When the
dot level is placed within the bias window, a resonance pattern
shows around $\epsilon_d$ with a peak developing in the accepting
$R$ reservoir and a corresponding dip showing at the $L$ side.
The dynamics shown in Fig. \ref{FigG1}(a) is fully coherent. Results presented
are before the recurrence time $\tau_{rec}^0\sim 2\pi/\Delta E$;
$\Delta E=2D/N$ is the mean spacing between energy levels \cite{Yukalov}.
At this time,  a complete depletion of certain levels occurs, and the dynamics is reversed.


We now wish to allow for elastic dephasing effects or inelastic interactions on the dot {\it only}.
We mimic such effects with ``probes", by augmenting the Hamiltonian (\ref{eq:H0}) by
an additional (finite size) electron reservoir $G$,
\bea H=H_0+H_G+V_G. \label{eq:H} \eea
Here $H_G=\sum_{g}\epsilon_gc_g^{\dagger}c_g$ and
$V_G=\sum_gv_gc_{g}^{\dagger}d$+h.c.. Inelastic effects are
introduced using a voltage probe \cite{Buttiker,Beenakker},
demanding that the net current from the dot to the $G$ unit
vanishes, $i_G=0$. This condition sets an effective chemical
potential for this reservoir. Alternatively, elastic dephasing
effects can be introduced using a dephasing probe, requiring that
$i_G(\epsilon)=0$, i.e., the charge current at a given energy
should vanish. Results are displayed in Fig. \ref{FigG1}(b) and (c):
The system approaches equilibrium (dephasing probe), and even
thermal equilibrium (voltage probe). These
results are  discussed in details below Eq. (\ref{eq:exact}). We now
explain how we time-evolved the system's density matrix under $H_0$
or $H$.


%


{\it QLE Method.} We aim at calculating the time evolution of {\it
all} two-body operators in the system. We begin with the trivial
part, the impurity (dot). Since it is coupled to many degrees of
freedom, its dynamics can be described using a quantum Langevin
equation \cite{QLE,Dhar,DharSC,DharP}. We review the steps involved,
so as to highlight the underling approximations, setting the limit
for the method applicability. We outline the derivation in the
absence of the $G$ probe; We generalize it later to
include such a device. In the Heisenberg representation the
operators satisfy the following equations of motion (EOM),
\begin{eqnarray}
\dot{c}_{d}&=&-i\epsilon_{d}c_{d}-i\sum_{l}v_{l}c_{l}-i\sum_{r}v_{r}c_{r}
\nonumber\\
\dot{c}_{l} &=&-i\epsilon_{l}c_{l}-iv_{l}c_{d},\,\,\,\,\,
\dot{c}_{r}=-i\epsilon_{r}c_{r}-iv_{r}c_{d}.
\label{eq:EOM}
\end{eqnarray}
Using a formal integration, the $l$ operators follow
\begin{eqnarray}
c_{l}(t)&=&e^{-i\epsilon_{l}(t-t_{0})}c_{l}(t_{0})-
iv_{l}\int_{t_{0}}^{t}d\tau
e^{-i\epsilon_{l}(t-\tau)}c_{d}(\tau)d\tau. \nonumber\\
\label{eq:clm}
\end{eqnarray}
Similar relations hold for the $R$ operators. We plug these expressions
into the dot EOM and get the exact result
%
\begin{eqnarray}
&&\dot{c}_{d}=-i\epsilon_{d}c_{d}
-\int_{t_{0}}^{t}d\tau\sum_{l}v_{l}^{2}
e^{-i\epsilon_{l}(t-\tau)}c_{d}(\tau)
\nonumber\\
&&
-\int_{t_{0}}^{t}d\tau\sum_{r}v_{r}^{2}e^{-i\epsilon_{r}(t-\tau)}c_{d}(\tau)
-i \eta^{L}(t) -i \eta^{R}(t).
 \label{eq:qle}
\end{eqnarray}
Here,
$\eta^{L}(t)\equiv\sum_{l}v_{l}e^{-i\epsilon_{l}(t-t_{0})}c_{l}(t_{0})$, and similarly
$\eta^{R}$, represent ``noise''.
We now assume that the second and third terms reduce, each, into a
dissipation term, further inducing an energy shift of the
dot energy, absorbed into the definition of $\epsilon_d$. This
is justified here as the metal grains play the role of charge
and energy baths with respect to the dot. Under the markovian
approximation, we reach the (time local) QLE
\bea \dot{c}_d&=&-i\epsilon_dc_d -i\eta^L(t) -i\eta^R(t) -
\Gamma(\epsilon_d) c_d, \label{eq:cd} \eea
with $\Gamma(\epsilon)=\sum_{\nu=L,R}\Gamma_{\nu}(\epsilon)$ and
e.g., $\Gamma_{L}(\epsilon)=\pi\sum_lv_l^2
\delta(\epsilon-\epsilon_l)$. Equation (\ref{eq:cd}) thus
relies on two assumptions:
(i)  a positive real self-energy function $\Gamma_{\nu}(\epsilon)$ can be written \cite{QLE}, and
(ii) the dot dynamics is slow relative to the reservoirs'
evolution.
We now use the exact equation (\ref{eq:clm}) and the reduced result (\ref{eq:cd}),
and derive analytic expressions for the expectation values $\langle
c_k^{\dagger}(t)c_j(t)\rangle\equiv {\rm
Tr}[\rho(t_0)c_k^{\dagger}(t)c_j(t)] $; $k,j=l,r,d$. Here
$\rho(t_0)=\rho_d\otimes \rho_L\otimes \rho_R$ is the factorized
time-zero density matrix of the system. The trace is
performed over all degrees of freedom. In particular, the following
initial condition is assumed
\bea
\langle c_d^{\dagger}(t_0)c_d(t_0)\rangle &=&0,\,\,\,
\langle c_l^{\dagger}(t_0)c_l(t_0)\rangle =f_L(\epsilon_l)\equiv f_l, \,\,\,
\nonumber\\
\langle c_r^{\dagger}(t_0)c_r(t_0)\rangle&=&f_R(\epsilon_r)\equiv f_r.
\eea
with $f_{\nu}(\epsilon)=[e^{\beta(\epsilon-\mu_{\nu})}+1]^{-1}$.
The resolved occupation of the, e.g., $L$ reservoir, is given by three contributions,
\bea &&p(\epsilon_l)\equiv\langle c_l^{\dagger}(t)c_l(t)\rangle
=\langle c_l^{\dagger}(t_0)c_l(t_0)\rangle
\nonumber\\
&&+iv_le^{-i\epsilon_l(t-t_0)} \int_{t_0}^{t}
e^{i\epsilon_l(t-\tau)} \langle c_d^{\dagger}(\tau)c_l(t_0) \rangle
d\tau + c.c.
\nonumber\\
&&+ v_l^2 \int_{t_0}^{t}\int_{t_0}^td\tau_1 d\tau_2 \langle
c_{d}^{\dagger}(\tau_1)c_d(\tau_2)\rangle
e^{i\epsilon_l(t-\tau_1)}e^{-i\epsilon_l(t-\tau_2)}. \nonumber
\label{eq:pl} \eea
The first term accommodates the initial condition. The second
element (denoted by $F_2$) represents first order reservoir-dot
coupling processes. The last term ($F_3$) corroborates higher
order effects, including population transfer from the $r$ reservoir,
\begin{widetext}
\bea
F_{2}&=&-v_{l}^{2}f_{l}\times (t-t_{0})\frac{2\Gamma}{\Gamma^{2}+
\epsilon_{dl}^{2}}-2v_{l}^{2}f_{l}\frac{\epsilon_{dl}^{2}
-\Gamma^{2}}{\left[\epsilon_{dl}^{2}+
\Gamma^{2}\right]^{2}}
+\frac{v_{l}^{2}f_{l}e^{-\Gamma(t-t_{0})}}{\left[\epsilon_{dl}^{2}+\Gamma^{2}\right]^{2}}\Biggl
\{2\left[\epsilon_{dl}^{2}-\Gamma^{2}\right]\cos\left[\epsilon_{dl}(t-t_{0})\right]+4\epsilon_{dl}\Gamma\sin\left[\epsilon_{dl}
(t-t_{0})\right]\Biggr\} \nonumber\\
F_{3}&=&v_{l}{}^{2}\sum_{k^{\prime}\in L,R}
\frac{v_{k^{\prime}}{}^{2}f_{k^{\prime}}}
{\Gamma^{2}+\epsilon_{dk^{\prime}}^{2}}\Biggl\{\frac{4\sin^{2}\left
[\frac{\epsilon_{lk^{\prime}}}{2}(t-t_{0})\right]}{\epsilon_{lk^{\prime}}^{2}}
+\frac{1}{\Gamma^{2}+\epsilon_{dl}^{2}}\left[e^{-2\Gamma(t-t_{0})}+1-e^{(t-t_{0})
(i\epsilon_{dl}-\Gamma)}-e^{-(t-t_{0})(i\epsilon_{dl}+\Gamma)}\right]\nonumber\\&+&\left[\frac{1-e^{-(t-t_{0})
(\Gamma+i\epsilon_{dl})}+e^{-(t-t_{0})(\Gamma+i\epsilon_{dk^{\prime}})}-e^{-i(t-t_{0})\epsilon_{lk^{\prime}}}}
{\left(\epsilon_{dl}-i\Gamma\right)\epsilon_{lk^{\prime}}}+c.c\right]\Biggr\}.
\label{eq:F2F3} \eea  \end{widetext}
Here, $\epsilon_{jk}=\epsilon_j-\epsilon_k$.
For simplicity, the energy dependence of the dissipation
terms is ignored. We can immediately confirm that
$i_{\nu}(t)=\frac{d}{dt}\sum_{k\in \nu}\langle
c_k^{\dagger}(t)c_k(t)\rangle$ produces the standard
expression for the charge current  \cite{Komnik}.
We have similarly derived closed  expressions for {\it all}
density matrix elements, including off-diagonal elements, e.g.,
$\langle c_l^{\dagger}(t)c_{l'}(t)\rangle$, ($l\neq l'$). These
lengthy expressions are not presented here.
Since the reservoirs include many states, after a
short time $\tau_t$, $\Gamma\tau_t \gtrsim2$, the dot dynamics
should remain fixed at a quasi steady-state value, up to the
recurrence time $\tau_{rec}^0$.
The interval $\tau_t<t<\tau_{rec}^0$ is identified as the quasi steady-state (QSS)
region

%


We now incorporate the $G$ reservoir, the probe,
into this QLE description: The dot dynamics follows Eq.
(\ref{eq:cd}) with an additional noise term $\eta^G$, and we
re-define the total hybridization,
$\Gamma=\Gamma_L+\Gamma_R+\Gamma_G$, $\Gamma_G(\epsilon)=\pi\sum_g
v_g^2\delta(\epsilon-\epsilon_g)$. As a result, all expectation
values follow a form technically identical to the $\Gamma_G=0$
limit. For example, the population $\langle
c^{\dagger}_l(t)c_l(t)\rangle$ obeys Eq. (\ref{eq:F2F3}), augmented
by $k'\in G$ terms in $F_3$.
%
%
To include
{\it inelastic scattering effects} of electrons on the dot,
we implement a voltage probe assuming
$f_G(\epsilon)=[e^{\beta(\epsilon-\mu_G)}+1]^{-1}$. The chemical
potential $\mu_G$ is set such that $i_G\equiv \frac{d}{dt}\sum_g{\langle
c_g^{\dagger}c_g\rangle }=0$ is satisfied at all simulation time. With
the motivation to explore situations beyond the linear response
regime \cite{DharSC}, we retrieve $\mu_G$ numerically, by employing
the Newton-Raphson method \cite{NR},
%
$\mu_G^{(k+1)}=\mu_G^{(k)}-i_G(\mu_G^{(k)})/i_G'(\mu_G^{(k)})$.
%
$\mu_G^{(0)}$ is the initial guess, $i'_G$ is the first derivative
with respect to $\mu_G$. In principle, one should adjust $\mu_G$
throughout the simulation, to eliminate population leakage from the
$L$-dot-$R$ system. However, we have practically found that within
the allowed simulation time (details below) we could safely assume a
QSS limit. The energy resolved charge current into $G$ can then be
written as
\bea
i_G(\epsilon)=
\frac{2\Gamma_G}{\pi}
\frac{\Gamma_R [f_G(\epsilon)-f_R(\epsilon)] +
\Gamma_L [f_G(\epsilon)-f_L(\epsilon)] }
{(\epsilon-\epsilon_d)^2+\Gamma^2},
\nonumber
\eea
%
with the total current
$i_G=\int  i_G(\epsilon) d\epsilon$.
{\it Quasi-elastic scattering} effects are implemented within a
dephasing probe, by demanding that $i_{G}(\epsilon)=0$, to yield
%
$f_{G}(\epsilon)=\frac{\Gamma_{R}f_{R}(\epsilon)+\Gamma_{L}f_{L}(\epsilon)}{\Gamma_{R}+\Gamma_{L}}$.
%

\begin{figure}[htbp]
\vspace{1mm} \hspace{0mm} {\hbox{\epsfxsize=60mm \epsffile{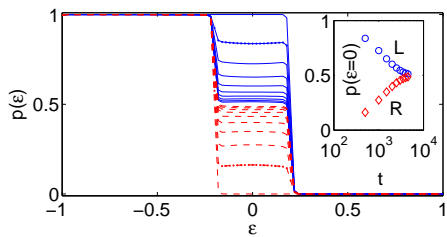}}}
\caption{Equilibration with a dephasing probe, using an
exact-unitary time evolution scheme, Eq. (\ref{eq:exact}).
$N_L=N_R=100$, $N_G=1500$, other parameters are the same as in Fig.
\ref{FigG1}(b). The $L$ (full) and $R$ (dashed) populations are
shown at $t=0:\delta t:9\delta t$,  $\delta t=500$.
The dotted lines mark the values reached before recurrence if $N_G=200$.
The inset demonstrates a slow-down in dynamics in approaching the
equilibrium state.} \label{FigE}
\end{figure}

{\it Exact method.} Even with the inclusion of the
probe, the QLE description is still technically limited
by the recurrence time $t<\tau_{rec}^0\propto N_{L,R}$ ($N_{L,R}$ is the number
of electronic states in a {\it single} metallic grain), though no
actual population recurrence does show in the dynamics. This is
because the validity of Eq. (\ref{eq:cd}) is limited by this time,
beyond which inter-reservoir recurrences, that may not show in the
overall behavior, take place. Using an {\it exact}, expensive, brute
force calculation, we can numerically simulate ($A\equiv
c_j^{\dagger}c_{k}$)
\bea
&&\langle A(t) \rangle = {\rm Tr}_B[\rho(t_0)e^{iHt}Ae^{-iHt}]
\nonumber\\
&&= {\rm lim}_{\lambda \rightarrow 0}\frac{\partial}{\partial
\lambda} {\rm Tr}_{B} [\rho_L\rho_R\rho_G\rho_d e^{iHt} e^{\lambda
A} e^{-iHt}], \label{eq:exact} \eea
using the fermionic trace formula \cite{Klich}. Here,
$\rho_{\nu}=e^{-\beta(H_{\nu}-\mu_{\nu}N_{\nu})}/Z_{\nu}$; $Z_{\nu}$
is the partition function. Such a calculation perfectly agrees with
QLE results, further confirming that beyond $\tau_{rec}^{0}$, where
 QLE description breaks down, the dynamics proceed towards
equilibrium, before $\tau_{rec}\propto \sum_{i=L,R,G}N_i$, see Fig. \ref{FigE}.
%
%
We now show an approach towards equilibrium within  $t<\tau_{rec}$
using the exact method, and $t<\tau_{rec}^{0}$ using the QLE technique.


\begin{figure}[htbp]
\vspace{1mm} \hspace{0mm} {\hbox{\epsfxsize=80mm \epsffile{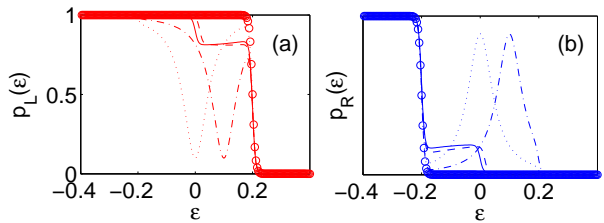}}}
\caption{(a) Occupation of $L$ bath at time $t=0$ ($\circ$) and
$t=1500$ for $\Gamma_G=0$ and $\epsilon_d=0$ (dotted), $\Gamma_G=0$
and $\epsilon_d=0.1$ (dashed-dotted), $\Gamma_G=0.4$ and
$\epsilon_d=0$ (full), $\Gamma_G=0.4$ and $\epsilon_d=0.1$ (dashed).
The latter two lines assume the voltage probe condition.
(b) Same for the $R$ side occupations.
Other parameters are the same as in Fig. \ref{FigG1}(c).
}
\label{FigG2}
\end{figure}


{\it Results.}
We identify thermal equilibration
in our peer-quantum system setup, by adjusting the conditions of Refs. \cite{Popescu,Integ2},
demanding that:
(i) The system should equilibrate, i.e., evolve towards some particular state, and stay close to it
for almost all time. Furthermore, the equilibrium state should be
(ii) independent of the dot properties-energetics
and initial state, (iii) insensitive to the precise initial state of each reservoir,
(iv) close to diagonal in the energy basis of its
eigen-Hamiltonian, and (v) a Gibbs state.

In Fig. \ref{FigG1}(b)-(c) we use the QLE method and follow the reservoirs' mutual
equilibration process, using either a
dephasing probe (b) or a voltage probe (c). We present the
reservoirs' occupation at times $t/75=0,10,20,100,200$. Beyond
$\tau_{rec}^0 \sim 1500$ the QLE description breaks down;
The data at later times has been generated by
restarting the simulation immediately before $\tau_{rec}^0$,
 using the $L$ and $R$ final diagonal-distribution as an
initial condition for a new run, with the $G$ bath re-adjusted to respect
the probe condition.
This process can be repeated, to
reach a complete equilibration. It can be realized experimentally by
fine-tweaking the voltage probe (introducing a dissipation mechanism into the dynamics) \cite{ExpVolt}.
%
When only dephasing effects are allowed, the system is approaching a
non-thermal state. Ultimately, the population of the two reservoirs
should reach a two-step function with
$p(\mu_R<\epsilon<\mu_L)\sim0.5$, since excess electrons at the $L$
side loose their phase memory, therefore, on average, half of them
populate the $R$ side in the long time limit. This
equilibrium state is sensitive to the precise details of the initial
electron distribution, as energy redistribution is not allowed. In
contrast, when inelastic effects are taking place, the system does
approach a Gibbs-like thermal state, a step function at zero
temperature.

Simulations with a dephasing probe using the exact-unitary method, Eq. (\ref{eq:exact}),
up to $\tau_{rec} \propto \sum_{i=L,R,G}N_i$,
are shown in Fig. \ref{FigE}, demonstrating a clear evolution towards an
equilibrium state.
We build a large
$G$ so as to delay recurrence. Results at earlier times do not depend on the size of $G$,
reinforcing the observation that $G$ acts as an agent in driving the $L$-$R$ mutual equilibration
process.
QLE data with $G$ bath tweaking nicely agrees with these results. 
In order to show the analogous behavior with a voltage probe, a dissipative mechanism should be
introduced into the $G$ bath, e.g., by building a hierarchy of its interactions with the $L$-$R$
system.

Fig. \ref{FigG2} proves that while under coherent evolution
the resonance peak emerges around the energy
$\epsilon_d$, in the presence of a voltage probe with (large enough)
$\Gamma_G$, the buildup of the equilibrium state systematically
occurs around the equilibrium Fermi energy, independently of the
link energetics. This holds even when the dot is placed {\it
outside} the bias window (not shown). Using a smaller value for
$\Gamma_G$, temporal features show, washed out gradually in
approaching the
equilibrium state.  Analogous trends also take place when allowing for dephasing only.

\begin{figure}[htbp]
\vspace{1mm} \hspace{0mm}
{\hbox{\epsfxsize=80mm \epsffile{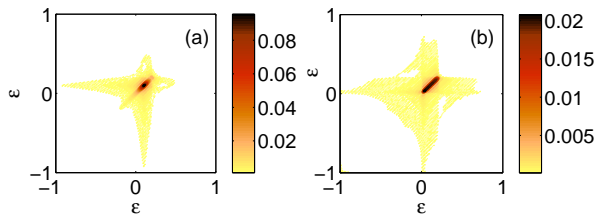}}} \caption{
Absolute values of the density matrix elements $\rho_{l,l'}$
at $t=1500$ (a) $\Gamma_G=0$, (b) using a voltage probe,
$\Gamma_G=0.4$.
Other parameters are the same as in Fig. \ref{FigG1}(c), $\epsilon_d=0.1$.
}
\label{FigDM}
\end{figure}

The thermal state should be diagonal in the energy eigenbasis of its
Hamiltonian \cite{Popescu}. In Fig. \ref{FigDM} we display the
density matrix (DM) $\rho_{l,l'}=\langle c_l^{\dagger}c_{l'}
\rangle$, excluding diagonals, with and without a voltage probe,
using the QLE technique.
This quantity is expected to oscillate in the long time limit since
the Hamiltonian is not diagonal in the (local) $l$ basis. We still
show the results in this basis, so as to manifest local
$\nu$-bath properties. A subtle source of complication 
is the fact that $\rho_{l,l'}$ decays (before $\tau_{rec}^0)$, even
without a probe, due to the finite-bias assumed as an initial
condition. One should therefore differentiate between bias-induced
and probe-induced decoherence processes. There are three significant
differences in the behavior of off-diagonal elements, with and
without the probe: (i) The absolute value of the coherences, at a
given time, is smaller when $\Gamma_G\neq 0$. (ii) The DM approaches
a diagonal form (strict diagonal values are not shown). (iii) When
$\Gamma_G=0$, oscillations occur around $\epsilon_d$. With the
probe, contributions appear mainly around the equilibrium Fermi
energy.


{\it Summary.} We provided evidence, using analytical and numerical tools,
 showing that two finite quantum systems,
coupled through a weak link where electrons can loose phase
memory or exchange energy, can equilibrate and even thermalize.
Dephasing and voltage probes are cunning techniques, allowing to
mimic memory loss and energy redistribution. We could follow the
time evolution of the peer metals numerically-exactly or using an
analytic QLE scheme, where tweaking the probes with a new initial
state drives the peer-metal system closer and closer to the
equilibrium state. Our results are significant for several reasons:
(i) We show that inelastic or dephasing effects on a very small- yet
essential- subset of the total system can drive the system towards a
global equilibrium state, where thermal equilibration is approached
in the former case. (ii) The colloquial, restrictive, ``nondegenerate
energy gap" condition \cite{Reimann,Short,Popescu,Hanggi} is not
assumed in our study. (iii) Standard QLE treatments \cite{QLE}
overlooked reservoirs' dynamics all-together. Here, we frame a new tool
for studying the dynamics of a system composed of many degrees of
freedom, by identifying a subsystem and resolving its dynamics, then using this
information backward, to re-explore the evolution of all degrees of
freedom. Future directions include the study of electron-electron
interaction effects \cite{IF}, and considering a quantum dot chain
between the two metal grains. Here, the charge current
coherent-diffusive crossover  \cite{Lebowitz} may reflect itself in
the energy reorganization of the reservoirs.

This work has been supported by NSERC. M.K. thanks Diptiman Sen for useful discussions.


\end{document}